\documentclass{article}

     \usepackage[preprint]{neurips_data_2022}

\usepackage[utf8]{inputenc} 
\usepackage[T1]{fontenc}    
\usepackage{hyperref}       
\usepackage{url}            
\usepackage{booktabs}       
\usepackage{amsfonts}       
\usepackage{nicefrac}       
\usepackage{microtype}      
\usepackage{xcolor}         
\usepackage{graphicx}
\usepackage{siunitx}
\usepackage{comment}
\usepackage{subcaption}

\newcommand{\posecheck}[1]{$\large{\text{P}}\small{\text{OSE}}\large{\text{C}}\small{\text{HECK}}$}

\setcitestyle{numbers,open={[},close={]},citesep={,}}

\title{Benchmarking Generated Poses: How Rational is Structure-based Drug Design with Generative Models?}

\author{%
  Charles Harris\thanks{Correspondance to \texttt{cch57@cam.ac.uk}} \\
  University of Cambridge\\
  \texttt{cch57@cam.ac.uk} \\
    \And
    Kieran Didi \\
    University of Cambridge \\
    \texttt{ked48@cam.ac.uk} \\
  \And
 Arian R. Jamasb \\
University of Cambridge\\
  \texttt{arj39@cam.ac.uk} \\
    \And
    Chaitanya K. Joshi \\
    University of Cambridge \\
    \texttt{ckj24@cam.ac.uk} \\
    \And
    Simon V. Mathis \\
    University of Cambridge \\
    \texttt{svm34@cam.ac.uk} \\
    \AND
    Pietro Lio \\
    University of Cambridge \\
    \texttt{pl219@cam.ac.uk}
    \And
    Tom L. Blundell \\
    University of Cambridge \\
   \texttt{tlb20@cam.ac.uk} \\
}

\begin{document}

\maketitle

\begin{abstract}
Deep generative models for structure-based drug design (SBDD), where molecule generation is conditioned on a 3D protein pocket, have received considerable interest in recent years. These methods offer the promise of higher-quality molecule generation by explicitly modelling the 3D interaction between a potential drug and a protein receptor. However, previous work has primarily focused on the quality of the generated molecules themselves, with limited evaluation of the 3D molecule \emph{poses} that these methods produce, with most work simply discarding the generated pose and only reporting a ``corrected” pose after after redocking with traditional methods. Little is known about whether generated molecules satisfy known physical constraints for binding and the extent to which redocking alters the generated interactions. We introduce \posecheck{}, an extensive analysis of multiple state-of-the-art methods and find that generated molecules have significantly more physical violations and fewer key interactions compared to baselines, calling into question the implicit assumption that providing rich 3D structure information improves molecule complementarity. We make recommendations for future research tackling identified failure modes and hope our benchmark can serve as a springboard for future SBDD generative modelling work to have a real-world impact. 
\end{abstract}

\section{Introduction}

Structure-based drug design (SBDD) \cite{blundell1996SBDD1, ferreira2015SBDD2, anderson2003SBDD3} is a cornerstone of drug discovery. It uses the 3D structures of target proteins as a guide to designing small molecule therapeutics. The intricate atomic interactions between proteins and their ligands shed light on the molecular motifs influencing binding affinity, selectivity, and drug-like properties. Employing computational methods such as molecular docking \cite{trott2010autodock, alhossary2015quickvina2}, molecular dynamics simulations \cite{klepeis2009MD}, and free energy calculations \cite{chipot2007free}, SBDD aids in the identification and optimization of potential drug candidates.

Deep generative models for SBDD have recently attracted considerable attention in the ML community \cite{du2022molgensurvey, isert2023structure}. These models learn from vast compound databases to generate novel chemical structures with drug-like properties \cite{gomez2018automatic}. By explicitly integrating protein structure information, these models aim to generate ligands with a higher likelihood of binding to the target protein. In particular, advancements in geometric deep learning \cite{bronstein2017gdl1, atz2021gdl2, jing2020learning} have led to a new suite of generative methods, enabling the design of 3D molecules directly within the confines of the target protein \cite{masuda2020LiGAN, luo20213dsbdd, peng2022pocket2mol, guan2023targetdiff, schneuing2022diffsbdd}. These methods, which concurrently generate a molecular graph and 3D coordinates, provide the significant advantage of obviating the need for determining the 3D pose \textit{post hoc} through traditionally slow molecular docking programs -- at least in theory.

\begin{figure}[t]
    \centering
    \includegraphics[width=\textwidth]{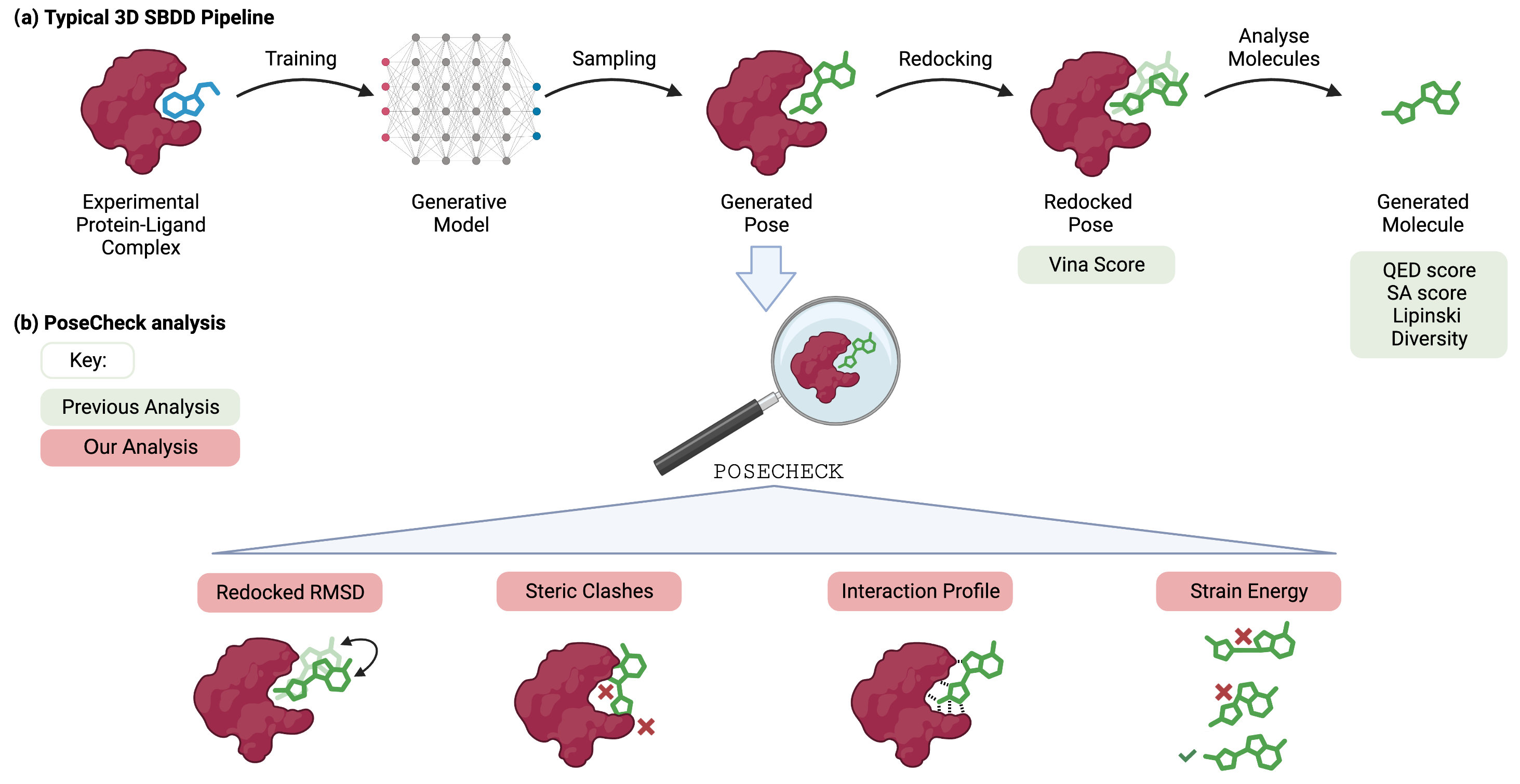}
    
    \caption{ \textbf{Top:} Overview of a conventional pipeline of SBDD with 3D generative modelling. A generative model is usually trained using experimental or synthetic protein-ligand complexes, from which new molecules and poses can be sampled \textit{de novo}. Typically, generated poses are discarded and redocked into the receptor, and primarily evaluated on 2D molecular graphs (e.g. QED). The effect of redocking on the final complex is often unknown, preventing understanding of the common failure modes of the trained model and therefore inhibiting progress.
    \textbf{Bottom:} the \posecheck{} benchmarks for generated poses include pipeline-wide as well as component-wise metrics, enabling a targeted evaluation of each model component guiding further model development.
    }
    \label{fig:main_figure}
    \end{figure}

Assessing the quality of molecules generated by these methodologies is not straightforward, with little work on experimental validation, especially for \textit{de novo} design \cite{baillif2023deep}. Typical evaluation metrics (Figure \ref{fig:main_figure}a) focus primarily on the 2D graph of the generated molecules themselves, measuring their physicochemical properties (e.g. QED \cite{bickerton2012QED}) and adherence to drug discovery heuristics (e.g. Lipinski's Rule of Five \cite{lipinski2012experimental}). For effective SBDD, we argue that it's equally essential to assess the quality of the generated \emph{binding poses} and their capacity to satisfy known biophysical prerequisites for binding (Figure \ref{fig:main_figure}b). This perspective is essential if these methods are to serve as practical alternatives to traditional virtual screening approaches in SBDD.

We hypothesise that multiple failure modes, undetected by currently applied metrics, are inherent within these methods. The situation is further complicated by the common practice of disregarding the initially generated pose and then redocking the molecule to attain a potentially enhanced pose. This strategy tends to focus on presenting only the outcomes of the redocked molecule, making the accurate assessment of pose quality an increasingly intricate challenge.


Our primary contributions are summarized as follows: We introduce \posecheck{}, a set of new biophysical benchmarks for SBDD models, expanding the traditional `pipeline-wide' framework by integrating `component-wise' metrics (i.e. generated and redocked poses), leading to comprehensive and precise model assessment. Utilizing this new framework, we evaluate a selection of high-performing machine learning SBDD methods, revealing two key findings: (1) generated molecules and poses often contain nonphysical features such as steric clashes, hydrogen placement issues, and high strain energies, and (2) redocking masks many of these failure modes. Based on these evaluations, we propose targeted recommendations to rectify the identified shortcomings. Our work thus provides a roadmap for addressing critical issues in SBDD generative modelling, informing future research efforts.

\vspace{10pt}

\section{Background}

\paragraph{Deep Generative Models for 3D Structure-based Drug Design} 

Many works have recently tried to recast the SBDD problem as learning the 3D conditional probability of generating molecules given a receptor, allowing users to sample new molecules completely \textit{de novo} inside a pocket. 
Common methods utilize Variational AutoEncoders (VAEs) \cite{kingma2013vaes}, Generative Adversarial Networks (GANs) \cite{goodfellow2014GANs}, Autoregressive (AR) models and recently Denoising Diffusion Probabilistic Models (DDPMs) \cite{ho2020ddpms}. LiGAN \cite{masuda2020LiGAN} uses a 3D convolutional neural network combined with a VAE model and GAN-style training. 3DSBDD \cite{luo20213dsbdd} introduced an autoregressive (AR) model that iteratively samples from an atom probability field (parameterised by a Graph Neural Network) to construct a whole molecule, with an auxiliary network deciding when to terminal generation. Pocket2Mol \cite{peng2022pocket2mol} extended this work with a more efficient sampling algorithm and better encoder. DiffSBDD \cite{guan2023targetdiff} and TargetDiff \cite{guan2023targetdiff} are both conditional DDPMs conditioned on the 3D target structure. All methods extract a point cloud from the end of the generative process and then draw bonds based on the distances between generated atoms using OpenBabel \cite{o2011openbabel}.

\paragraph{Related work}

\citet{guan2023targetdiff} perform limited analysis of small chemical sub-features, such as agreement to experimental atom-atom distances and the correctness of aromatic rings within the generated molecule. \citet{baillif2023deep} emphasize the necessity of 3D benchmarks for 3D generative models. However, both of these works study the molecules in isolation rather than the protein-ligand context.

\section{Methods}

In order to evaluate the quality of generated poses and their capacity to facilitate high-affinity protein-ligand interactions, we present a variety of computational methods and benchmarks in this section. These methodologies provide a thorough perspective on the poses produced and illuminate the ability of generative models to generate trustworthy and significant ligand conformations. Full implementation details are given in Appendix A.

\paragraph{Interaction fingerprinting}

Interaction fingerprinting is a computational method utilized in SBDD to represent and analyze the interactions between a ligand and its target protein. This approach encodes specific molecular interactions, such as hydrogen bonding and hydrophobic contacts, in a compact and easily comparable format -- typically as a bit vector, known as a \emph{interaction fingerprint} \cite{bouysset2021prolif, marcou2007optimizing}. Each element in the interaction fingerprint corresponds to a particular type of interaction between the ligand and a specific residue within the protein binding pocket. In a similar way to chemical similarity \cite{willett1998chemical}, encoding these interactions enables rapid comparison of different ligands or binding poses by computing the Tanimoto similarity between fingerprints:

\[
T(A, B) = \frac{{|A \cap B|}}{{|A| + |B| - |A \cap B|}}
\]

where $A$ and $B$ are sets, and $|A|$ represents the cardinality (size) of set $A$. We compute interactions using the ProLIF library \cite{bouysset2021prolif}.

\paragraph{Steric clashes}

In the context of molecular interactions, the term \emph{steric clash} is used when two neutral atoms come into closer proximity than the combined extent of their van der Waals radii \cite{ramachandran2011clash1}. This event indicates an energetically unfavourable \cite{buonfiglio2015protein}, and thus physically implausible, interaction. The presence of such a clash often points towards the current conformation of the ligand within the protein being less than optimal, suggesting possible inadequacies in the pose design or a fundamental incompatibility in the overall molecular topology. Hence, the total number of clashes serves as a vital performance metric in the realm of SBDD. 
We stipulate a clash to occur when the pairwise distance between a protein and ligand atom falls below the sum of their van der Waals radii, allowing a clash tolerance of 0.5 Å.

\paragraph{Strain-energy}

Strain energy refers to the internal energy stored within a ligand as a result of conformational changes upon binding. When a ligand binds to a protein, both the ligand and the protein may undergo conformational adjustments to accommodate each other, leading to changes in their bond lengths, bond angles, and torsional angles. These changes can cause strain within the molecules, which can affect the overall binding affinity and stability of the protein-ligand complex \cite{perola2004conformational}.
Whilst there is always a trade-off between enthalpy and entropy, generally speaking, lower strain energy results in more favourable binding interactions and potentially more effective therapeutics.
We calculate the strain energy as the difference between the internal energy of a relaxed pose and the generated pose (without pocket). Both relaxation and energy evaluation are computed using the Universal Force Field (UFF) \cite{rappe1992uff} using RDKit. 





\paragraph{Evaluation by redocking}

Our final assessment involves measuring the level of agreement between the docking programs and the molecules produced by the learned distribution in the generative model. Although this is the most coarse-grained approach we employ and docking programs come with their inherent limitations, they nevertheless contain useful proxies and serve as valuable tools for comparison. In this procedure, we redock the generated pose into the pocket using QuickVina2 \cite{alhossary2015quickvina2}. Following this, we compute the Root Mean Squared Deviation (RMSD) between the generated pose and the docking-predicted one across all generated molecules, thereby obtaining a distribution of RMSD values.

\section{Results}

\begin{figure}[t]
    \centering
    \includegraphics[width=\textwidth]{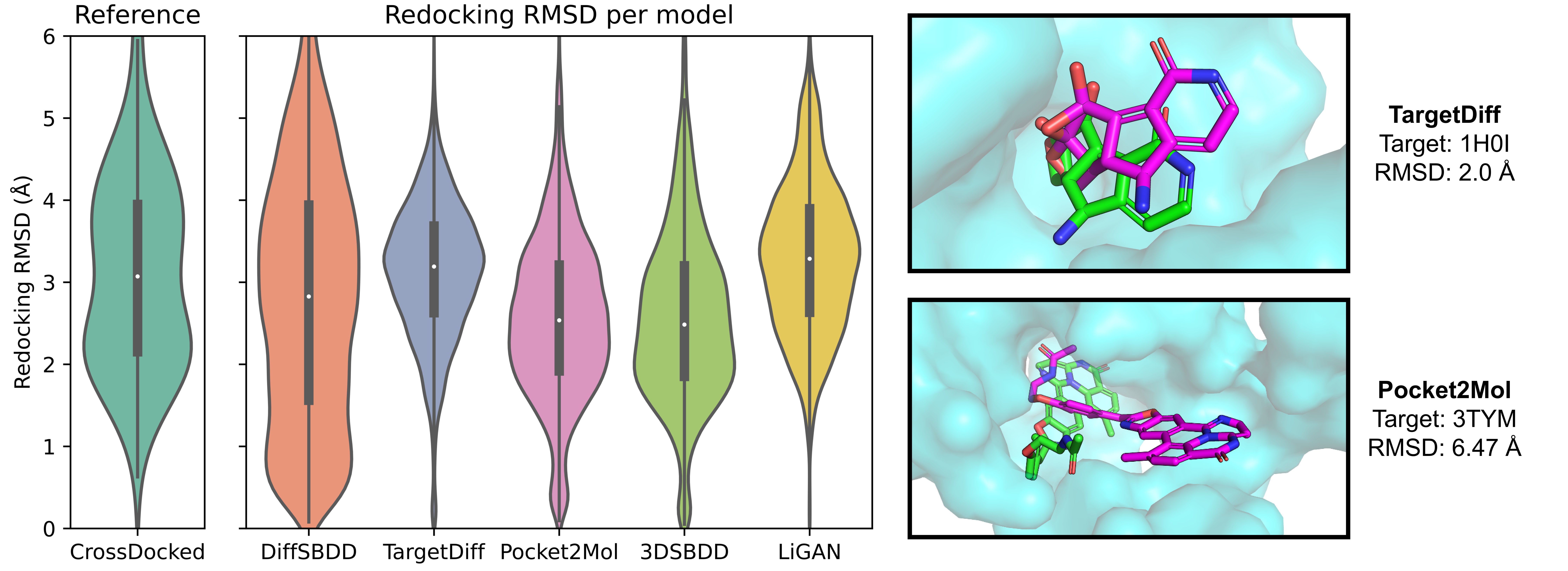}
    \caption{RSMD between the generated and redocked poses using the popular docking framework Vina. 
    \textbf{Left:} Violin plots of all redocking RMSDs. CrossDocked violin is where we have redocking the poses from the original dataset.
    \textbf{Right:} Illustrative examples (generated in magenta, redocked in green). We observe that the generated and redocked poses often differ significantly. The top figure shows an example of RMSD~2\AA, which already results in a clearly different pose. Failure modes such as entire flips of a generated molecule upon re-docking are not uncommon, indicating that the generated pose is often in disagreement with the principles behind common docking software.}
    \label{fig:rmsd_plots}
\end{figure}

\subsection{Experimental Setup}

In our study, we evaluate the quality of poses from five recent methods: LiGAN \cite{masuda2020LiGAN}, 3DSBDD \cite{luo20213dsbdd}, Pocket2Mol \cite{peng2022pocket2mol}, TargetDiff \cite{guan2023targetdiff} and DiffSBDD \cite{schneuing2022diffsbdd}. All models were trained on the CrossDocked2020 \cite{francoeur2020crossdocked} dataset using the dataset splits computed in \citet{peng2022pocket2mol}, which used a train/test split of 30\% sequence identity to give a test set of 100 target protein-ligand complexes which we use for evaluation. For each model, we sampled 100 molecules per target. 
During inference, the model is given a reduced PDB file containing only the atoms for a single pocket within the test set, so there is no element of blind docking during generation or subsequent redocking\footnote{Note illustrative figures may show full proteins.}.

\subsection{Agreement with docking functions}

\paragraph{Results}


In an effort to discern the impact of redocking within 3D generative SBDD post-processing pipelines, our preliminary action involves determining the extent to which the redocked pose preserves accurate information from the initially generated binding mode. Therefore, we proceed to measure the RMSD between the pose our model has generated and the pose that has been redocked using QuickVina2 \cite{alhossary2015quickvina2}. A lower RMSD value would denote a higher degree of agreement between the generated binding modes and the scoring function. To provide perspective, it's worth noting that a carbon-carbon bond generally measures 1.54 \AA \space in length.

The density distributions of various methods are illustrated in Figure \ref{fig:rmsd_plots}. All methods exhibit remarkably high average redocking RMSD scores, with no method achieving a median below 2 \AA. Interestingly, the diffusion methods perform worse than the AR (Auto-Regression) methods, as evidenced by DiffSBDD and TargetDiff achieving median scores of 2.83 and 3.19 \AA, respectively, while Pocket2Mol and 3DSBDD achieve 2.57 and 2.83 \AA, respectively. LiGAN performs similarly to the diffusion methods, with a median score of 3.27. It is worth noting that DiffSBDD shows a significantly wider range of scores compared to all other methods, with a standard deviation of 1.54\AA, whereas other methods have values ranging from 0.83 to 1.04 \AA. Full extended results for all section are given in Appendix B.

\paragraph{Discussion}

\begin{figure}[t]
    \centering
    \includegraphics[width=\textwidth]{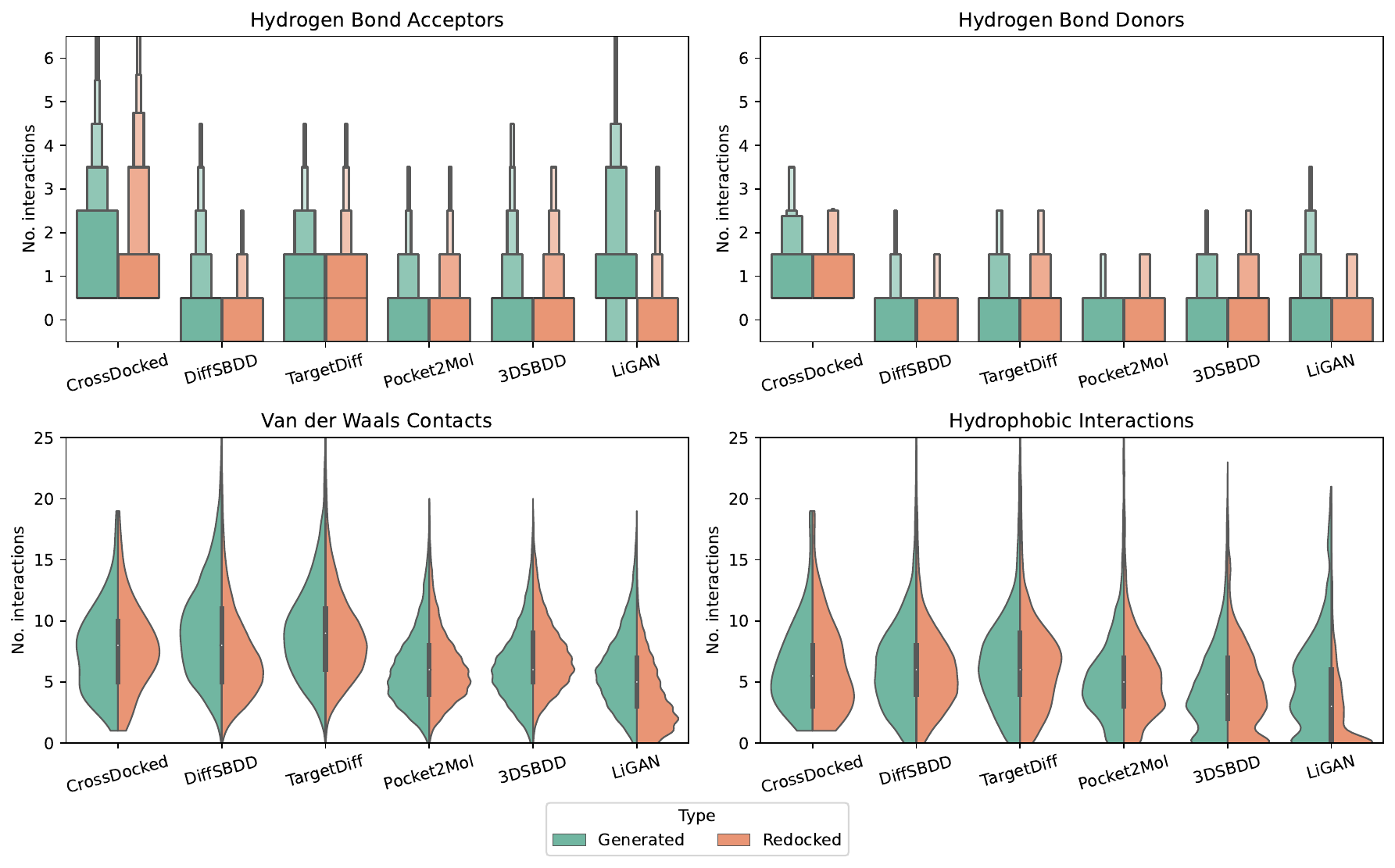}
    \caption{Interactions between protein and ligands as seen in reference datasets (green) and generated molecules (orange). The frequency of (a) hydrogen bond acceptors, (b) hydrogen bond donors, (c) Van der Waals contacts and (d) hydrophobic interactions are considered. We find that generative models have significant trouble making hydrogen bond interactions compared to baseline.}
    \label{fig:interactions}
\end{figure}

These findings raise concerns for several reasons. They expose the minimal concordance between the binding models learned by these methods and the established QuickVina2 methodology \cite{alhossary2015quickvina2}. More critically, they underline the lack of accurate evaluations of generative models' capability to produce realistic binding poses; instead, these models tend to generate drug-like molecules with vague binding modes, later rectified through docking. Simultaneously, we should consider that the similar performance observed in the RMSD distribution (median 3.07 \AA RMSD) when redocking the CrossDocked test set using QuickVina2 might suggest two potential scenarios. The CrossDocked dataset could be plagued by quality issues, or this could reflect a variance in the docking programs used - SMINA is utilized by CrossDocked \cite{koes2013lessons}, whereas QuickVina2 is favoured by the broader ML community \cite{luo20213dsbdd, peng2022pocket2mol, guan2023targetdiff, schneuing2022diffsbdd}.

\paragraph{Limitations}

Furthermore, it may be that all methods have learnt excellent binding models (or at least as good as smina \cite{koes2013lessons}), but that we cannot resolve this due to the use of QuickVina2 \cite{alhossary2015quickvina2} as the scoring function. Future work could consider performing this benchmarking using the same docking protocol as CrossDocked \cite{francoeur2020crossdocked} or higher-fidelity docking programs such as Glide \cite{friesner2004glide}. However, we think this further underlines the issues of the currently used evaluation paradigm.


\subsection{Protein-ligand interaction analysis}

\begin{figure}[t]
    \centering
    \includegraphics[width=\textwidth]{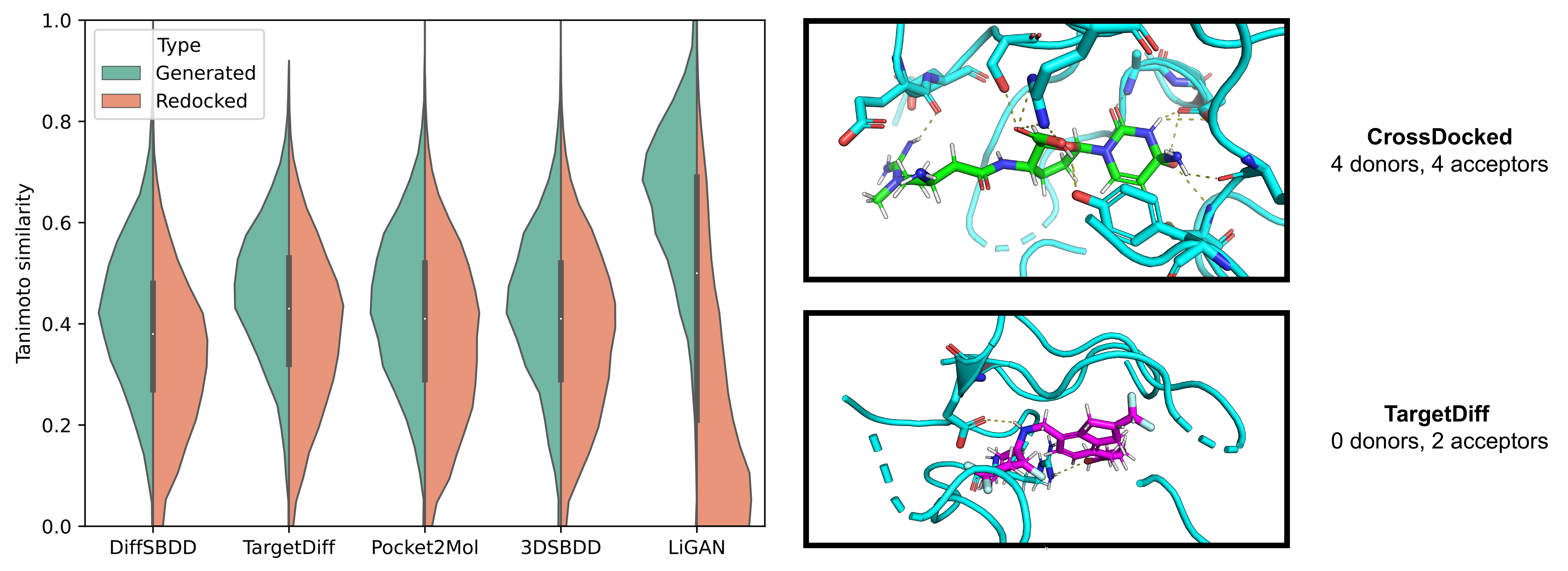}
    \caption{\textbf{Left:} Interaction similarity between the generated molecules/interactions and the CrossDocked test set molecules. \textbf{Right:} examples of a molecule from the test set (green) and a generated pose from TargetDiff (magenta).\vspace{-15pt}}
    \label{fig:interaction_similarties}
\end{figure}


\paragraph{Evaluation}

Below describe the classes of interaction that we evaluate.
\textbf{Hydrogen bonds} (HBs) are a type of interaction that occurs between a hydrogen atom that is bonded to a highly electronegative atom, such as nitrogen, oxygen, or fluorine \cite{pimentel1971hydrogenbonds}. They are key to many protein-ligand interactions \cite{chen2016hydrogenbond_protein} and require very specific geometries to be formed \cite{brown1976geometry_hbonds}. The directionality of HBs confers unique identities upon the participating atoms: hydrogen atoms attached to electronegative elements are deemed `donors', whilst the atom accepting the HB is termed an `acceptor'. 
\textbf{Van der Waals contacts} (vdWs) are interactions that occur between atoms that are not bonded to each other. These forces can be attractive or repulsive and are typically quite weak \cite{andersson1998van}. However, they can be significant when many atoms are involved, as is typical in protein-ligand binding \cite{barratt2005van_protein}.
\textbf{Hydrophobic interactions} are non-covalent interactions that occur between non-polar molecules or parts of molecules in a water-based environment. They are driven by the tendency of water molecules to form hydrogen bonds with each other, which leads to the exclusion of non-polar substances. This exclusion principle prompts these non-polar regions to orient away from the aqueous environment and towards each other \cite{meyer2006hydrophobic}, thereby facilitating the association between protein and ligand molecules \cite{patil2010hydrophobic_binding}.


\paragraph{Results}

Distributions of various kinds of interactions are shown in Figure \ref{fig:interactions}. We consider whether our generative models can design molecules with adequate hydrogen bonding and find that no method can match or exceed the baseline. In the reference set, CrossDocked, the mode of HBs for both acceptors and donors is 1, with means of 2.23 and 1.66 for acceptors and donors respectively. Strikingly, we find that in all generated poses for all models (except LiGAN HB acceptors), the \emph{most common number of HB acceptors and donors is 0}, with means varying between 0.68-1.73 for HB acceptors and 0.39-0.85 for HB donors. We find an average difference of 0.50 and 0.81 HBs between the best-performing models and the baseline for acceptors and donors respectively. Results for Van der Waals contacts and hydrophobic interactions are closer to the dataset baseline.

\paragraph{Discussion}

Conventional wisdom would suggest that many minor imperfections in the generated pose would be simply fixed by redocking the molecule (e.g. moving an oxygen atom slightly to complete a hydrogen bond.) We find this is in fact rarely the case, with redocking sometimes being significantly deleterious (see examples of LiGAN in Figure \ref{fig:interactions}), suggesting that there are either limitations in the docking function used or, more likely, the generated interaction was physically implausible to begin with.



\paragraph{Interaction similarities to reference poses}

We also consider the interaction similarities between the generated poses/interactions and the baseline test set in order to study generated binding modes (see Figure \ref{fig:interaction_similarties}). Here, we take the bit vectors constructed for Figure \ref{fig:interactions} and compute the Tanimoto distance between the reference pose and all generated poses. We observe that all methods perform similarly (both generated and redocked) except for LiGAN, which achieves the highest similarity before being completely degenerated by redocking, suggesting their model has done well to maximise placing atoms where they form interactions with the pocket, but the poses themselves are not meaningful. 

\subsection{Clash scores}

\begin{figure}[h]
    \centering
    \includegraphics[width=0.9\textwidth]{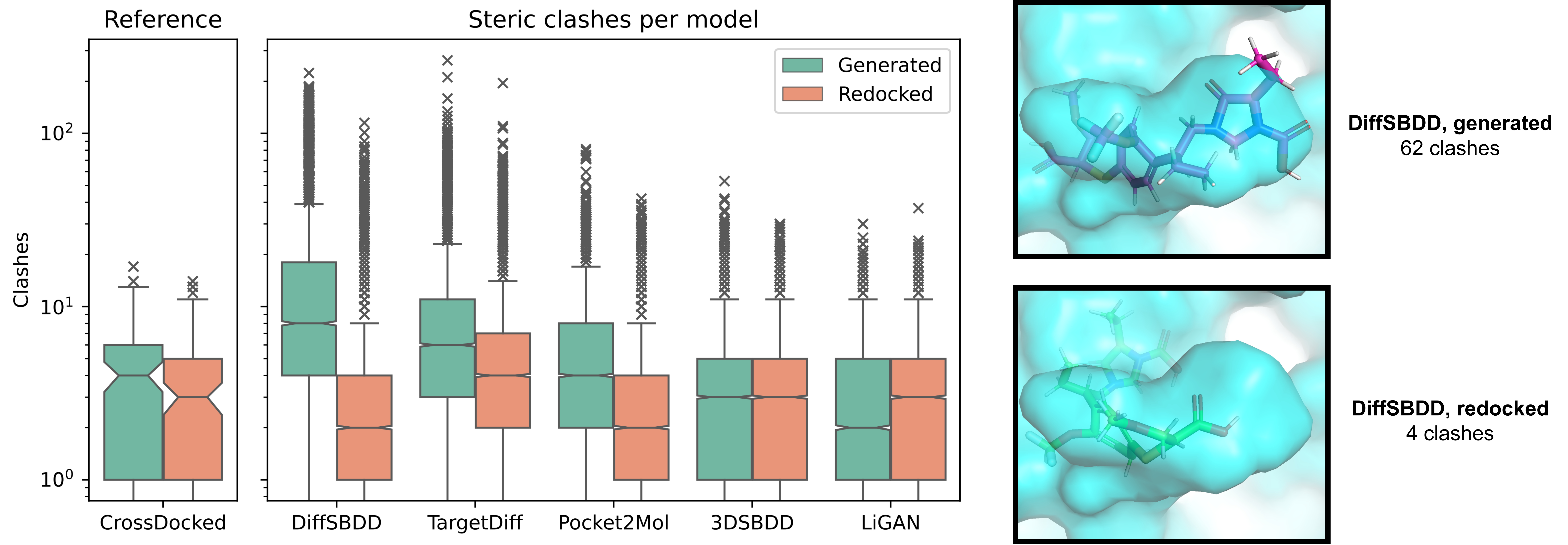}
    \caption{Clash scores.
    \vspace{-15pt}}
    \label{fig:clashes}
\end{figure}

\paragraph{Results}

Figure \ref{fig:clashes} presents the results of the steric clash analysis. In summary, the latest methods, particularly those employing diffusion models, exhibit poor performance in terms of steric clashes compared to the baseline, with a significant number of outliers. Although redocking mitigates clashes to some extent, it does not always resolve the most severe cases.

The CrossDocked test set has a low number of clashes with few extreme examples, with a mean of 4.59, upper quantile of 6 and maximum value of 17. In terms of generated poses, the older methods perform best, with 3DSBDD and LiGAN having means of 3.79 and 3.40 clashes respectively. Pocket2Mol, an extension of 3DSBDD, performs worse with a mean clash score of 5.62 and upper quantile of 8 clashes. Finally, the diffusion-based approaches perform the worst with mean clash scores of 9.08 and 15.33 for TargetDiff and DiffSBDD respectively. The tail end of their distributions is also high, with both methods having upper quantiles of 11 and 18 respectively, with TargetDiff having the worst case of 264 steric clashes. Redocking the molecules generally fixed many clashes and improved scores. The mean clash score for Pocket2Mol improves from 5.62 to 2.98, TargetDiff from 9.08 to 5.79 and DiffSBDD from 15.34 to 3.61. 

\paragraph{Discussion}

Interestingly, DiffSBDD and TargetDiff, which are considered state-of-the-art based on mean docking score evaluations \cite{guan2023targetdiff, schneuing2022diffsbdd}, exhibit subpar performance in their number of clashes. They aim to learn atom position distributions without explicit constraints on final placements. While DiffSBDD starts with a performance deficit, its enhanced clash mitigation during redocking elevates its results to match the baseline, highlighting methodological distinctions between it and TargetDiff. Notably, 3DSBDD and LiGAN show low clash scores, with the former positioning atoms within a predefined voxel grid \cite{luo20213dsbdd} and the latter applying a clash loss \cite{masuda2020LiGAN}.

Our findings affirm the assumption that redocking alleviates many minor clashes, akin to the force-field relaxation step in AlphaFold2 \cite{jumper2021alphafold}. We initially speculated that molecules with clashes exceeding 100 had been mistakenly generated inside the protein pocket. Yet, we often discovered fragments within highly constrained nooks, especially worsened with the addition of hydrogen atoms.

\paragraph{Limitations}

An important consideration to bear in mind is that proteins are not entirely rigid receptors. They can often experience limited conformational rearrangements to accommodate molecules of varying shapes and sizes \cite{davis1999rigid_receptor}. Consequently, conducting generation and redocking in a rigid receptor environment may not yield accurate scores for potentially plausible molecules.
Note all these results are with a \emph{generous} clash tolerance of 0.5 \AA \space (roughly half the ddW radii of a hydrogen atom), in order to be able to resolve differences between methods.

\subsection{Strain energy}
\label{sec:strain}

\begin{figure}[h]
    \centering
    \includegraphics[width=\textwidth]{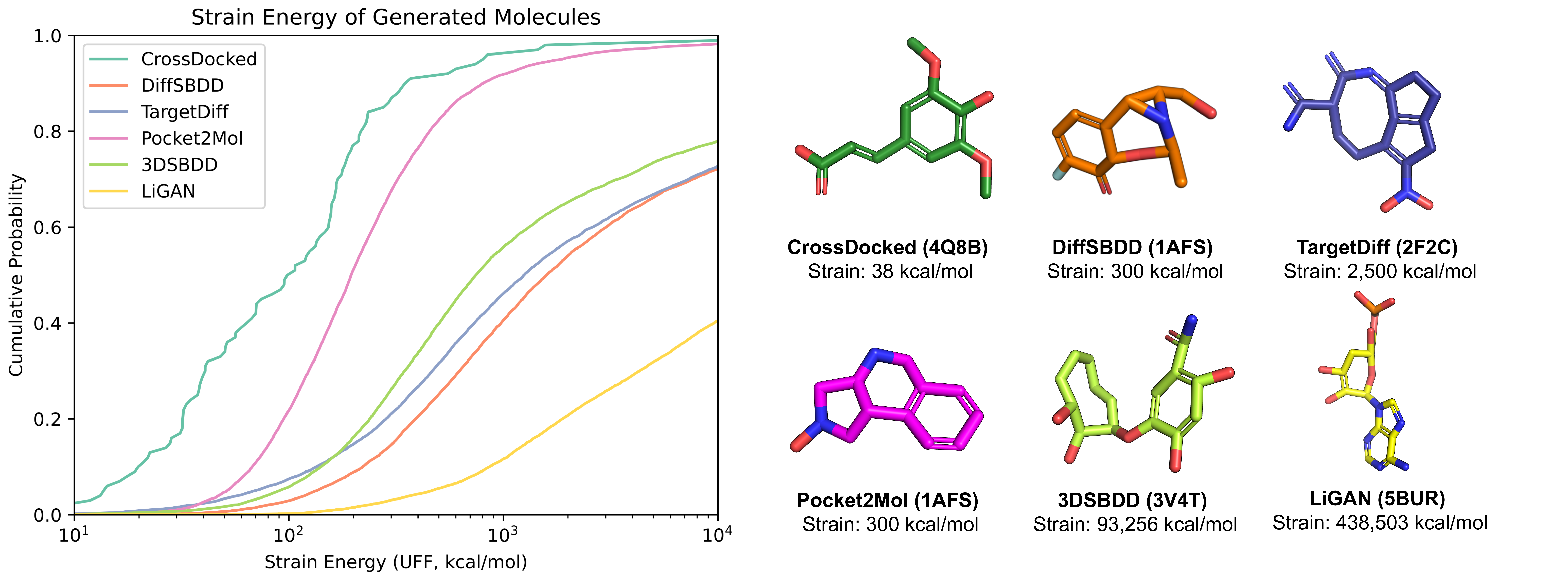}
    \caption{\textbf{Left:} CDF of strain energies. \textbf{Right:} Examples of molecules with high strain energy.}
    \label{fig:strain_energy}
\end{figure}

\paragraph{Results}

To conclude our study, we provide an analysis of the strain energy \cite{perola2004conformational} of the generated poses. Force field relaxation before docking is a common post-processing step of many generative SBDD pipelines, masking some potential issues with the generated geometries less clear. This allows us to evaluate the generated molecules for things like unrealistic bond distance or impossible geometries in rings.

Figure \ref{fig:strain_energy} displays the cumulative density function (CDF) of strain energy for the generated molecules, with the CrossDocked dataset serving as a baseline (Note: the x-axis is on a logarithmic scale). We focus on median values in our discussion since they are more representative in this context due to the presence of extreme outliers, with \emph{mean} values ranging from approximately $10^4$ to $10^{15}$ kcal/mol. None of the generative methods yields molecules exhibiting strain energy close to that of the test set, which has a median strain energy of 102.5 kcal/mol.

\paragraph{Discussion}

Intriguingly, both  of the diffusion-based methodologies (DiffSBDD and TargetDiff) perform similarly poorly, reporting median values of 1243.1 and 1241.7 kcal/mol, respectively. This could suggest issues with the currently used noised schedules \cite{chen2023importance} of these methods for ultra-precise atom position refinement (discussed in Section \ref{sec:recommendations}). 3DSBDD performs to the same order of magnitude, with a median strain energy of 592.2 kcal/mol, suggesting that placing atoms into a discretized voxel space \cite{luo20213dsbdd}, while good for avoiding clashes, has a detrimental impact on the strain energy. 

Pocket2Mol performs by far the best in terms of strain energy, with a median of 194.9 kcal/mol. The method provides perhaps the finest-grained control over exact coordinates generated, by first choosing a focal atom and then generating a new atom coordinate directly using an equivariant neural network \cite{jing2020learning, peng2022pocket2mol}, which may allow for more precise placement. LiGAN exhibits the highest strain energy, with a median value of 18693.8 kcal/mol, indicating the poorest performance in this context.

\paragraph{Limitations}

The exceedingly high strain energy values observed in this scenario should be approached with considerable prudence. For comparison, the combustion of TNT releases approximately 815 kcal/mol. \cite{rinkenbach1930tnt}. This data is not to be perceived as absolute, but rather illustrative of the extent to which our generated geometries deviate markedly from the standard distribution for the force field. This further underscores the existing issues. It is also conceivable that these poses might not even be initialized within more sophisticated, high-fidelity force fields \cite{brooks2009charmm}.


\vspace{-8pt}

\section{Recommendations for future work}
\label{sec:recommendations}

\paragraph{Exploring reduced-noise sampling strategies}

Interestingly, both diffusion-based works (DiffSBDD and TargetDiff) performed similarly in terms of strain energy (see Section \ref{sec:strain}). We hypothesize this may be due to the injection of random noise into the coordinate features at all but the last step of stochastic gradient Langevin dynamics samplings \cite{welling2011bayesian}, making it challenging to construct precise bond angles etc. Here, inspiration could be taken from protein design. For example, Chroma develops a low-temperature sampling regime to reduce the effect of noise \cite{ingraham2022chroma}, FrameDiff effectively scales down injected noise \cite{yim2023framediff}, both resulting in a substantial increase in sample quality with an acceptable decrease in sample diversity. 

\paragraph{Heavily penalise steric clashes during training}

All evaluated methods frequently create steric clashes, resulting in physically unrealizable samples. We suggest that mitigating steric clashes is key for the next generation of SBDD models. This could be done via extra loss terms, for example, by including a distogram loss as in AlphaFold2 \cite{jumper2021alphafold} or the steric clash loss in LiGAN \cite{masuda2020LiGAN}. A similar loss-based approach has been effective in mitigating chain-breaks diffusion models for protein backbone design \cite{yim2023framediff}.

\paragraph{Consider representing hydrogens}

Virtually all work in ML for structural biology chooses to not explicitly represent hydrogen atoms \cite{jumper2021alphafold, peng2022pocket2mol, luo20213dsbdd, yim2023framediff, schneuing2022diffsbdd, guan2023targetdiff}, under the assumption that they can be \emph{implicitly} learned and reasoned over with deep neural networks. However, our analysis of hydrogen bond networks within generated molecules found that generative methods struggle to handle the precise geometries required to make a hydrogen bond \cite{brown1976geometry_hbonds} (even when redocked). Despite the increased computational cost, we therefore recommend that future work explores their inclusion.



\section{Conclusion}


In conclusion, this work presents a comprehensive exploration of structure-based drug design (SBDD) methodologies with deep generative models. We advocate for the need to consider \emph{both} the quality of the generated molecules \emph{and} the quality of the binding poses in these models, calling for an expanded evaluation of SBDD. The application of deep generative models in SBDD holds promise for developing innovative drug-like molecules.
However, for SBDD approaches to realise that potential, we must establish a rigorous evaluation regimen of both the generated molecules and their interaction with the target -- as proposed in this paper. Our research provides a solid evaluation regimen for future advancements in this field and we hope that it stimulates further development towards more efficient drug discovery processes.




\section{Acknowledgements}

The authors would like to thank Arne Schneuing and Yuanqi Du for their discussions and help refining the ideas presented here.

\section{Code Availability}

We will make the code available for \posecheck{} in due course.



%







\bibliography{references}

\newpage

\appendix

\section{Extended Implementation}

All generative methods accessed were trained using the same dataset and splits as proposed in \citet{peng2022pocket2mol}. For all methods, docking was performed using QuickVina2. For generated poses, we sourced molecules from \citet{schneuing2022diffsbdd} for DiffSBDD, and \citet{guan2023targetdiff} for CrossDocked, TargetDiff, Pocket2Mol, 3DSBDD and LiGAN (where they also performed redocking). We will provide an updated version of this manuscript where we perform all redocking using entirely our own pipeline.

\section{Extended results}

\begin{figure}[t]
    \centering
    \includegraphics[width=\textwidth]{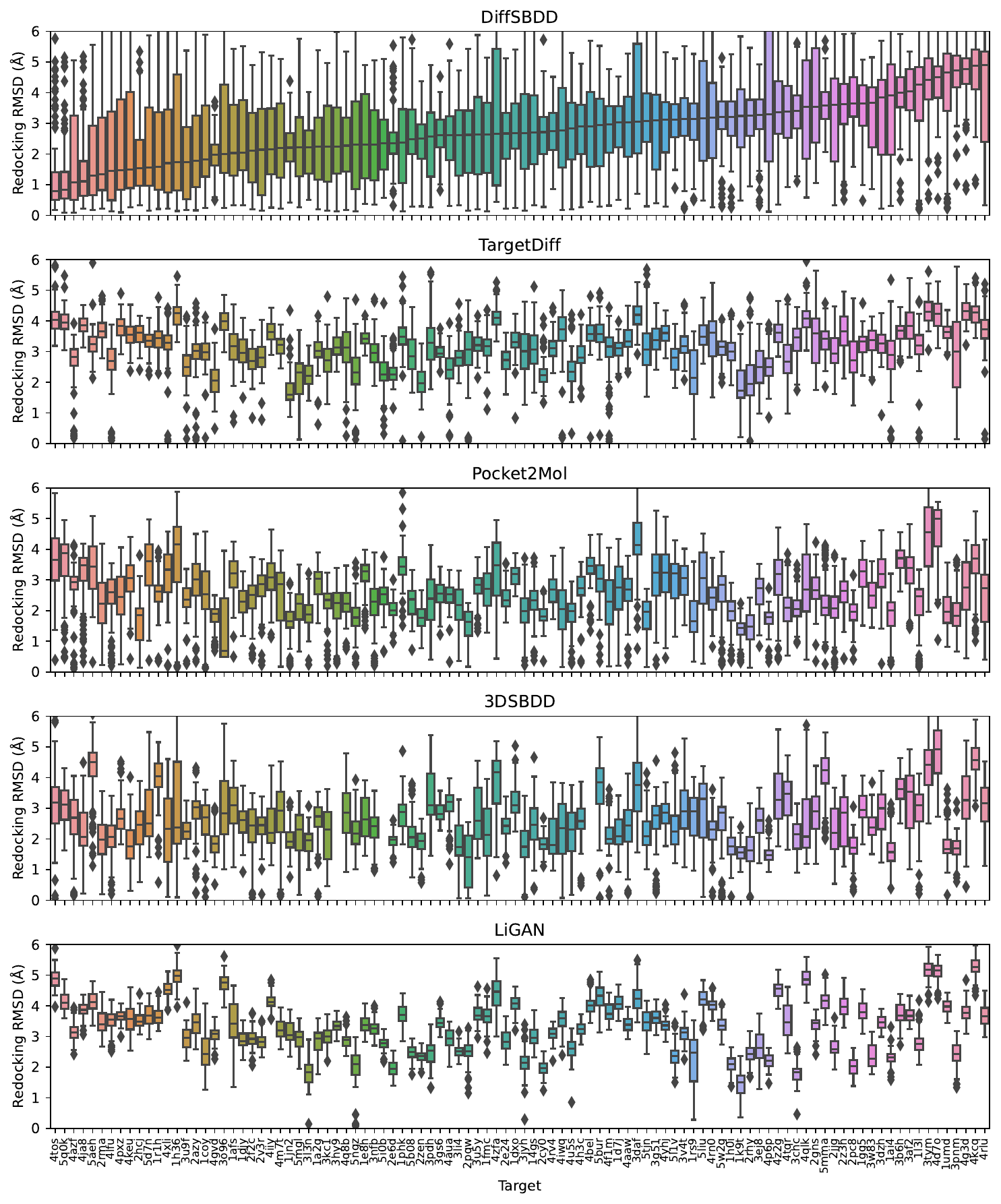}
    \caption{Redocking RMSD per method per target for CrossDocked test set. Order is determined arbitrarily by median score per target for DiffSBDD.
    }
    \label{fig:rmsd_per_target}
\end{figure}

\begin{figure}[t]
    \centering
    \includegraphics[width=\textwidth]{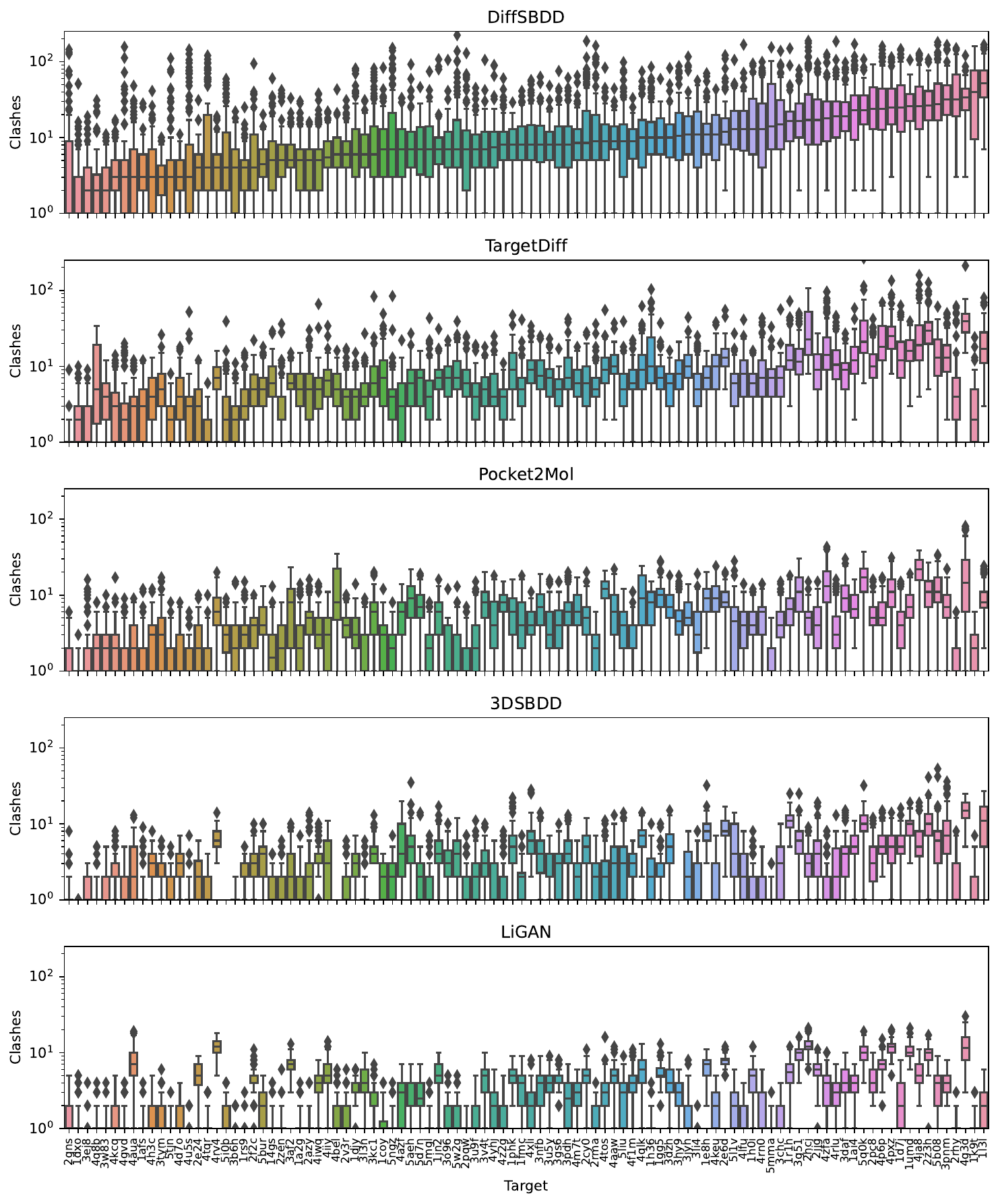}
    \caption{Steric clashes per method per target for generated poses in the  CrossDocked test set. Order is determined arbitrarily by median score per target for DiffSBDD.
    }
    \label{fig:clashes_per_target_generated}
\end{figure}

\begin{figure}[t]
    \centering
    \includegraphics[width=\textwidth]{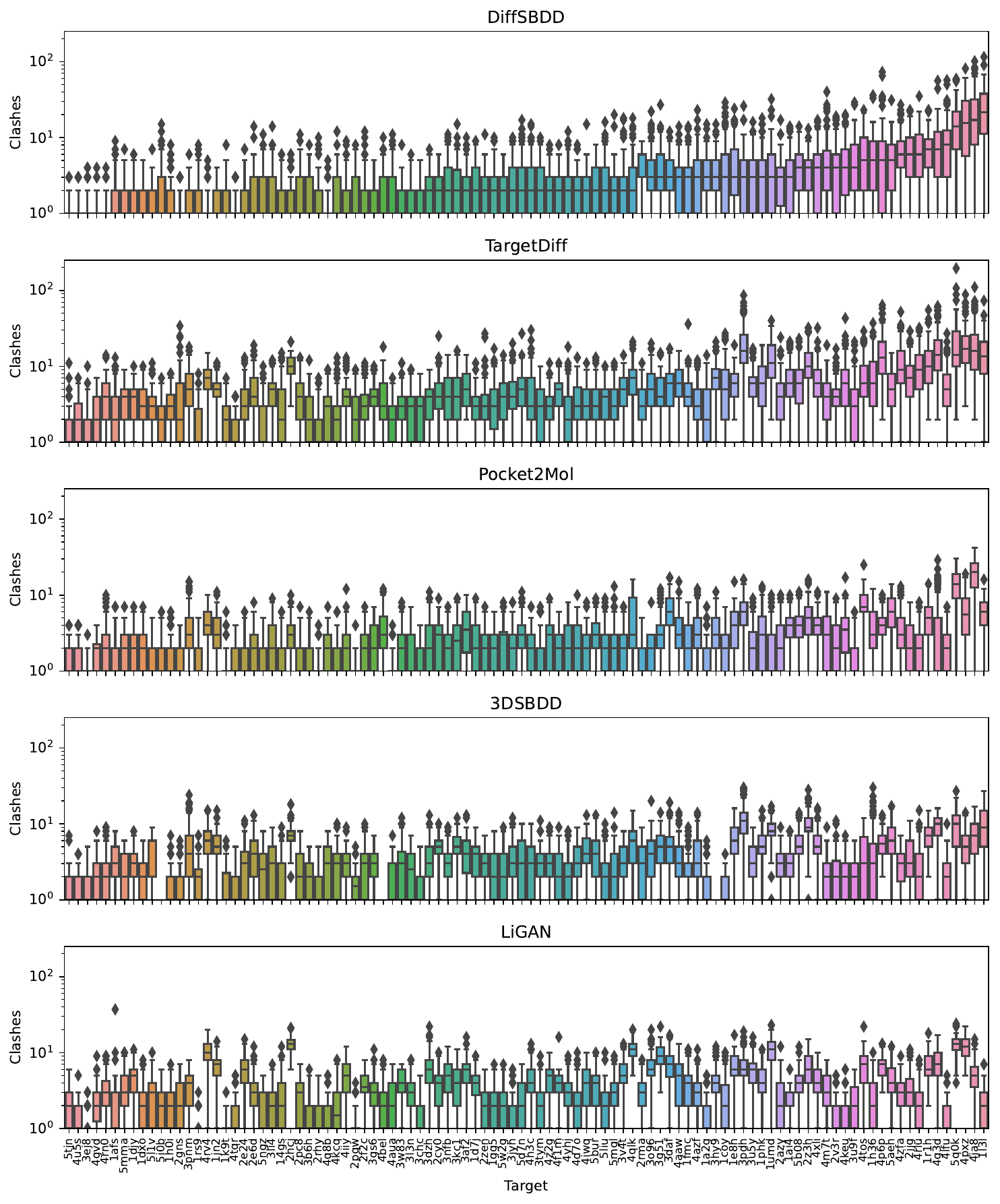}
    \caption{Steric clashes per method per target for redocked poses in the CrossDocked test set. Order is determined arbitrarily by median score per target for DiffSBDD.
    }
    \label{fig:clashes_per_target_docked}
\end{figure}

In Figure \ref{fig:rmsd_per_target}, we provide the per target redocking RMSDs per method. Figure \ref{fig:clashes_per_target_generated} and \ref{fig:clashes_per_target_docked} show the number of steric clashes per target for the generated and redocked poses respectively.

\end{document}